\documentclass[aip,jcp,reprint]{revtex4-1}
\usepackage{epsfig}
\usepackage{graphicx}
\usepackage{amsmath}
\usepackage{amssymb}
\usepackage{amsfonts}  
\usepackage{fancyhdr}
\usepackage{bm}
\usepackage{dcolumn}
\usepackage{subfigure}
\begin{document}

\title{Embedded density functional theory for covalently bonded and strongly interacting subsystems}

\author{Jason D. Goodpaster}
\author{Taylor A. Barnes}
\author{Thomas F. Miller, III$^{*}$}
%\email{tfm@caltech.edu}

% Use once for each address, use symbols \dag \ddag \S \P $\|$
% to connect authors with addresses

\affiliation{
Division of Chemistry and Chemical Engineering, California Institute of Technology, Pasadena, CA 91125
}
\date{\today}

\begin{abstract}

Embedded density functional theory (e-DFT) 
 is used to describe the electronic structure of 
strongly interacting molecular subsystems. 
We present a general implementation of the Exact Embedding (EE) method [J. Chem. Phys. \textbf{133}, 084103 (2010)] 
to calculate the large contributions of the non-additive kinetic potential (NAKP) in such applications.
Potential energy curves are computed for the dissociation of  Li$^+$-Be, CH$_3$-CF$_3$, and hydrogen-bonded water clusters, 
and e-DFT results obtained using the EE method are compared with those obtained using approximate kinetic energy functionals.
In all cases, the EE method preserves excellent agreement with reference Kohn-Sham calculations, whereas the approximate functionals lead to qualitative failures in the calculated energies and equilibrium structures. 
We also demonstrate an accurate pairwise approximation to the NAKP that 
allows for efficient parallelization of the EE method in large systems; benchmark calculations on molecular crystals reveal ideal, size-independent scaling of wall-clock time with increasing system size. 
\end{abstract}
\maketitle

\section{Introduction}

Important methodological challenges in electronic structure theory include the long-timescale simulation of  \emph{ab initio} molecular dynamics and the seamless combination of high- and low-level electronic structure methods in complex systems.
Methods that exploit the intrinsic locality of molecular interactions have demonstrated encouraging progress towards these goals.\cite{Li06,Li09,Fed07,Col06,Cha04,Cor91,wesorig, carterembedding,Wes06,Yan91,Lia03,Wer03,Zha03,Jac07,Sau89,Pac92,Jug04,Ell10}

In particular, orbital-free embedded density functional theory (e-DFT) offers a formally exact 
approach to electronic structure theory in which the interactions between subsystems are evaluated in terms of their electronic densities. \cite{Cor91,wesorig, carterembedding,Wes06}
Recent work 
has demonstrated that constructing the embedded subsystems from individual molecules leads to a linear-scaling electronic structure approach that maps naturally onto distributed-memory parallel computers, \cite{Ian06,Jac07} and it provides a systematic framework for calculating electronic excited states in condensed phase systems.\cite{Kam10, Vis08b}
However, approximate treatments of the non-additive kinetic potential (NAKP) limit the accuracy of this approach in applications involving strongly interacting subsystems.\cite{Vis09}  For example, severe artifacts in the structure of liquid water, including the complete absence of a second peak in the oxygen-oxygen radial distribution function, have been predicted from existing approximations to the NAKP,\cite{Ian06} and e-DFT applications involving covalently bonded embedded subsystems have also been shown to qualitatively fail.\cite{Vis09, Vis10b, Rei08}
The development of improved methods to address the NAKP problem will open new doors for on-the-fly, massively parallel electronic structure calculations in general, condensed-phase systems. 

In this paper, we describe progress towards the development of accurate, scalable treatments for the NAKP in e-DFT.  We provide the first molecular applications of our recently developed Exact Embedding (EE) method, \cite{tfm10jason} demonstrating that it successfully describes the breaking of covalent bonds and hydrogen bonds with chemical accuracy. 
Additionally, 
we introduce and numerically demonstrate a pairwise approximation to the NAKP, which allows for the scalable implementation of the EE method in large systems.  
Benchmark calculations are presented for systems with up to 125 molecules, demonstrating that parallel implementation of the method enables constant system-size scaling of the wall-clock calculation time.  

\section{Theory}

\subsection{Orbital-free Embedded DFT}

We utilize the orbital-free e-DFT formulation of Cortona \cite{Cor91} and Wesolowski and coworkers. \cite{wesorig,Wes06} 
For the case in which the total electronic density $\rho_\textrm{AB}$ is partitioned into two subsystems, % contributions, 
$\rho_\textrm{AB}=\rho_\textrm{A}+\rho_\textrm{B}$, the corresponding one-electron orbitals obey the Kohn-Sham Equations with Constrained Electron Density (KSCED),\cite{Wes06} 
\begin{eqnarray}
\left[-\frac12\nabla^2+v_\textrm{eff}[\rho_\textrm{A},\rho_\textrm{AB};{\bf r}]\right]\phi_i^\textrm{A}({\bf r})=\epsilon_i^\textrm{A}\phi_i^\textrm{A}({\bf r})  \label{kscedA}\\
\left[-\frac12\nabla^2+v_\textrm{eff}[\rho_\textrm{B},\rho_\textrm{AB};{\bf r}]\right]\phi_j^\textrm{B}({\bf r})=\epsilon_j^\textrm{B}\phi_j^\textrm{B}({\bf r}), \label{kscedB}
\end{eqnarray}
where $i=1,\ldots,N^\textrm{A}$, $j=1,\ldots,N^\textrm{B}$, and 
$N^\textrm{A}$ and $N^\textrm{B}$ are the number of electrons in the respective subsystems. $v_\textrm{eff}$ is the  effective potential for the coupled one-electron equations, such that
\begin{eqnarray}
v_\textrm{eff}[\rho_\textrm{A},\rho_\textrm{AB};{\bf r}]&=& v_{\textrm{ne}}(\mathbf{r})
+ v_{\textrm{J}}[\rho_\textrm{AB}; \mathbf{r}]+v_{\textrm{xc}}[\rho_\textrm{AB}; \mathbf{r}]  \nonumber\\
&&+\ v_{\textrm{nad}}[\rho_{\textrm{A}},\rho_{\textrm{AB}}; \mathbf{r}],
\label{vsced_orig}
\end{eqnarray}
where  the $N_\textrm{nuc}$ nuclei occupy positions $\{{\bf R}_i\}$,
\begin{eqnarray}
v_{\textrm{ne}}(\mathbf{r}) &=& -\sum_{i}^{N_\textrm{nuc}}\frac{Z_{i}}{|{\bf r}-{\mathbf{R}}_{i}|}, \\
v_{\textrm{J}}[\rho; \mathbf{r}]&=&\int \frac{\rho({\bf r}')}{|{\bf r}'-{\bf r}|}d{\bf r}',\\
v_{\textrm{xc}}[\rho; \mathbf{r}]&=&\left[\frac{\delta E_\textrm{xc}[\rho]}{\delta\rho}\right]\!\!({\bf r}),
\end{eqnarray}
 and $E_\textrm{xc}[\rho]$ is the exchange-correlation functional.
 $v_{\textrm{nad}}[\rho_{\textrm{A}},\rho_{\textrm{AB}}; \mathbf{r}]$ is the potential due to the  non-additive kinetic energy for non-interacting electrons,
 such that
 \begin{equation}
 v_{\textrm{nad}}[\rho_{\textrm{A}},\rho_{\textrm{AB}}; \mathbf{r}]  = \left[ \frac{\delta{T_{\textrm{s}}^{\textrm{nad}}{[\rho_{\textrm{A}},\rho_{\textrm{B}}]}}} {\delta{\rho_{\textrm{A}}}} \right] \!\!(\mathbf{r}),
 \label{nakp_def}
 \end{equation}
 where  $T_{\textrm{s}}^\textrm{nad}[\rho_\textrm{A},\rho_\textrm{B}]\equiv T_{\textrm{s}}[\rho_\textrm{AB}]-T_{\textrm{s}}[\rho_\textrm{A}]-T_{\textrm{s}}[\rho_\textrm{B}]$.  The subsystem densities are constructed from the corresponding KS orbitals, using $\rho_\textrm{A}({\bf r}) = \sum_{i=1}^{N^\textrm{A}} |\phi^\textrm{A}_i({\bf r})|^2$ and $\rho_\textrm{B}({\bf r}) = \sum_{j=1}^{N^\textrm{B}} |\phi^\textrm{B}_j({\bf r})|^2$.  
Eqs.~\ref{kscedA}-\ref{nakp_def} are easily generalized for the e-DFT description of multiple embedded subsystems. \cite{Cor91,Ian06}

Two aspects of e-DFT are worth emphasizing. Firstly, like conventional Kohn-Sham density functional theory (KS-DFT), it is a theory that is exact in principle, but practical calculations must employ approximations to the unknown exchange-correlation functional.  Secondly, unlike conventional KS-DFT, the embedding formulation introduces the NAKP, $ v_{\textrm{nad}}[\rho_{\textrm{A}},\rho_{\textrm{AB}}; \mathbf{r}] $, since the %one-electron 
orbitals for subsystem A are not necessarily orthogonal to those of subsystem B.  Without knowledge of the exact functional for the non-interacting kinetic energy, this creates a second source of approximation in e-DFT calculations.  The significance of the NAKP is system-dependent, with the most severe cases including those for which the subsystem densities %$\rho_{\textrm{A}}$ and $\rho_{\textrm{B}}$ 
greatly overlap; no approximate kinetic energy functional has been previously demonstrated to yield accurate results for embedded subsystems that are connected by covalent bonds. \cite{Wes06, Vis08, Vis09, Vis10, Rei08}

  \subsection{Exact Calculations of NAKP}
  
We have recently developed the Exact Embedding (EE) method to calculate the NAKP in the e-DFT framework.\cite{tfm10jason} 
   The general method can be summarized for two embedded subsystems as follows: 
A Levy constrained search (LCS)\cite{Lev79} %(LCS)
   or equivalent technique is first used to determine the full set of orthogonal KS orbitals, \{$\phi_i^{\textrm{AB}}$\}, that correspond to the total density $\rho_{\textrm{AB}}$  from the latest iteration of Eqs. \ref{kscedA}-\ref{vsced_orig}. % and  \ref{kscedB}.
  Then, from the KS orbitals $\{\phi_i^{\textrm{AB}}\}$, $\{\phi_i^{\textrm{A}}\}$, and $\{\phi_i^{\textrm{B}}\}$, the corresponding kinetic potentials are calculated using the exact result of King and Handy, \cite{Kin00}
\begin{equation}
v_{T_{\textrm{s}}}(\textbf{r}) = \frac{\sum_{i=1}^n(-\frac{1}{2}\phi_i(\textbf{r})\nabla^2 \phi_i(\textbf{r})) - \epsilon_i\phi_i(\textbf{r})^2)}{\rho(\textbf{r})} + \mu,
\label{king_handy}
\end{equation}
where $n$ is the number of occupied orbitals, $\epsilon_i$ is the KS eigenvalue corresponding to orbital $\phi_i$, and $\mu$ is a constant.
Finally, the NAKP needed for the next iteration of Eqs. \ref{kscedA}-\ref{vsced_orig} is calculated directly from the difference
\begin{equation}
v_{\textrm{nad}}[\rho_{\textrm{A}},\rho_{\textrm{AB}}; \mathbf{r}] = v^\textrm{AB}_{T_{\textrm{s}}}(\textbf{r})-v^\textrm{A}_{T_{\textrm{s}}}(\textbf{r}),
\label{NAKP}
\end{equation}
where the superscripts %on the right-hand side %(RHS) of 
in this equation indicate the orbital set to which each kinetic potential corresponds.

Rather than explicitly performing the LCS, %which is challenging from a computational perspective, 
we use the equivalent protocol of Zhao, Morrison, and Parr (ZMP)\cite{Zha92,Zha93,Zha94} to obtain the 
exact non-interacting kinetic energy and the KS orbitals \{$\phi_i^{\textrm{AB}}$\}. %for the total system.   
This requires solution of the following one-electron equations
\begin{equation}
\left[-\frac12\nabla^2+v_\textrm{ext}({\bf r})+ v^{\lambda}_{\textrm{c}}(\textbf{r})\right]\phi_{i,\lambda}^\textrm{AB}({\bf r})=\epsilon_{i,\lambda}^\textrm{AB}\phi_{i,\lambda}^\textrm{AB}({\bf r})
\label{ZMP1e}
\end{equation}
\noindent
in the limit $\lambda\rightarrow\infty$, where $i=1,\ldots, (N^{\textrm{A}}+N^{\textrm{B}})$, and
\begin{equation}
v^{\lambda}_{\textrm{c}}(\textbf{r}) = \lambda \int \frac{\rho(\textbf{r}')-\rho_\textrm{AB}(\textbf{r})}{\left|\textbf{r}'-\textbf{r}\right|}d\textbf{r}'.
\label{ZMP}
\end{equation}
$v_\textrm{ext}({\bf r})$ corresponds to any well-behaved external potential,\cite{Zha93,Zha94} and various choices for this potential are described in Sec.~III~B.  In practice, Eq.~\ref{ZMP1e} is solved at several large, finite values of $\lambda$, and the KS orbitals and eigenvalues, as well as the final non-interacting kinetic energy, are obtained via extrapolation. \cite{Zha92,Zha93,Zha94}
In Sec.~V, we discuss a technique to robustly implement the ZMP step for NAKP calculations in large systems.

The EE method outlined in Eqs. \ref{king_handy} - \ref{ZMP} is unique in that it allows for the formally exact calculation of the total electronic density within the e-DFT framework, using integer orbital occupancies and without approximations to the NAKP.
The method was previously demonstrated for atomic systems with strongly overlapping subsystem densities,\cite{tfm10jason} and the current paper presents its first molecular applications. 
We note that several other groups have also used density inversion techniques to calculate the NAKP, assuming that the total electron density is already available from another electronic structure calculation.\cite{Vis10b,Ron08,Ron09}
In particular, Visscher and coworkers have applied this approach to molecular systems with the aim of developing improved non-additive kinetic energy functionals.\cite{Vis10b}
Furthermore, Partition DFT has been introduced as a formally exact embedding scheme in which subsystem densities are described using partially occupied orbitals, and it has been applied to one-dimensional model systems.\cite{Ell10}

\section{Implementation Details}

We have implemented e-DFT in the Molpro quantum chemistry package,\cite{Molpro} allowing for calculation of the NAKP with either approximate functionals or the EE method. In this section, methodological and numerical aspects of the implementation are discussed. 

\subsection{Supermolecular vs.~Monomolecular Basis Sets}

The atom-centered basis sets used to solve the KSCED (Eqs. \ref{kscedA} and \ref{kscedB}) are implemented using two different conventions.\cite{Wes97, Vis09}  In the monomolecular basis set convention, the density for each embedded subsystem is described using only the basis functions that are centered on atoms belonging to that subsystem.  
In the supermolecular basis set convention, the density for each embedded subsystem is described using the same basis set, which includes functions that are centered on all atoms in the system.  The supermolecular basis set convention provides a closer approximation to the complete basis set limit, although it is more costly. 

\subsection{ZMP Step}

In our implementation, the ZMP step of the EE method is performed by solving Eq.~\ref{ZMP1e} for six large, finite values of $\lambda$. 
The KS orbitals $\{\phi_i^{\textrm{AB}}\}$ are then obtained from extrapolation of the atomic orbital coefficients for the $\{\phi_{i,\lambda}^{\textrm{AB}}\}$, using a third-order polynomial in $\lambda^{-1}$, and normalization of the extrapolated orbitals is enforced \emph{a posteriori}.
The KS eigenvalues $\{\epsilon_{i}^\textrm{AB}\}$ are similarly obtained from extrapolation of the $\{\epsilon_{i,\lambda}^\textrm{AB}\}$.
$T_{\textrm{s}}[\rho_\textrm{AB}]$ is calculated analytically from the extrapolated orbital coefficients, which ensures that the total energy from the EE method is bound from below by the KS-DFT energy.

In the limit $\lambda\rightarrow\infty$, the solutions to Eq.~\ref{ZMP1e} are independent of the choice of external potential $v_\textrm{ext}({\bf r})$,\cite{Zha92,Zha93,Zha94}
although $v_\textrm{ext}({\bf r})$ does affect the convergence with increasing $\lambda$. Various options where thus considered, including
\begin{eqnarray}
v_\textrm{ext}({\bf r}) &=& v_{\textrm{ne}}(\mathbf{r}),  \label{vext1} \\
v_\textrm{ext}({\bf r}) &=& v_{\textrm{ne}}(\mathbf{r}) + \left( 1 - \frac{1}{N^\textrm{A}+N^\textrm{B}} \right) v_\textrm{J} [ \rho_\textrm{AB}; {\bf r}],  \label{vext2}\\
v_\textrm{ext}({\bf r}) &=& v_{\textrm{ne}}(\mathbf{r})
+v_\textrm{J} [ \rho_\textrm{AB}; {\bf r}]  + v_\textrm{xc} [ \rho_\textrm{AB}; {\bf r}].
\label{vext3}
\end{eqnarray}
At every iteration of the KSCED, these versions of $v_\textrm{ext}({\bf r})$ are all available without the need for additional computation.  
Test calculations have indicated that the external potential in Eq.~\ref{vext3} leads to the fastest convergence of the extrapolation with increasing $\lambda$, and this potential is used in all results for the EE method reported in Sec.~IV.

\subsection{NAKP Numerics for Regions of Weak Density Overlap}

Numerical evaluation of  the kinetic potential from Eq.~\ref{king_handy}  is unstable in regions for which the corresponding density vanishes.  The problem is 
 exacerbated by the incorrect distance dependence of the low-density tails obtained from calculations using Gaussian-type orbitals (GTOs).\cite{Kin00} 
However, these numerically treacherous regions correspond to weak overlap between subsystem densities, where the magnitude of the NAKP is necessarily small and easily approximated.\cite{wesorig}
We thus utilize a density-based criterion to switch from the exact expression for the kinetic potential to a numerically stable approximation, such as the Thomas-Fermi (TF) kinetic potential. % in regions of weak density overlap.
The protocol used to perform this switching is described below.
	
In a first step, we calculate the constant shift that is needed to match the exact result for each kinetic potential to the corresponding TF result in a prescribed switching region.
 Specifically, for each of the kinetic potentials % the total kinetic potential and each subsystem kinetic potentials 
(i.e., $v_{T_{\textrm{s}}}(\textbf{r})\in\left\{ v^\textrm{AB}_{T_{\textrm{s}}}(\textbf{r}), v^\textrm{A}_{T_{\textrm{s}}}(\textbf{r}), v^\textrm{B}_{T_{\textrm{s}}}(\textbf{r})\right\}$ which correspond respectively to $\rho(\textbf{r})\in\left\{ \rho_\textrm{AB}(\textbf{r}), \rho_\textrm{A}(\textbf{r}), \rho_\textrm{B}(\textbf{r})\right\}$),
the average difference 
($\Delta\in\left\{ \Delta^\textrm{AB}, \Delta^\textrm{A}, \Delta^\textrm{B}\right\}$) 
between the results from Eq.~\ref{king_handy} and from the TF functional
 is evaluated in the vicinity of the $\rho(\textbf{r})=\rho'$ density isosurface.
Each $\Delta$ is
computed over 
% obtained from the weighted average of the difference between the exact and approximate expressions taken over
 gridpoints in the region $\xi < f [\rho; \textbf{r}] < (1-\xi)$, where 
\begin{equation}
f [\rho; \textbf{r}] = \frac{1}{ e^{\kappa(\rho(\textbf{r}) - \rho^\prime)} +1 },
\label{switch}
\end{equation}
$\xi$, $\kappa$, and $\rho'$ are parameters that define the switching region, and
the relative contribution from each gridpoint is weighted according to
\begin{equation}
\omega[\rho;\textbf{r}]  = e^{-\kappa(\rho_\textrm{AB}(\textbf{r}) - \rho(\textbf{r}))}.
\label{weights}
\end{equation}
Note that the weighting function in Eq.~\ref{weights} is uniform for the case of $\rho=\rho_\textrm{AB}$; for cases in which $\rho$ is one of the subsystem densities,
$\omega[\rho;\textbf{r}] $ preferentially selects values for which $\rho(\textbf{r})\approx\rho_\textrm{AB}(\textbf{r})$.

In a second step, each kinetic potential % $v_{T_{\textrm{s}}}(\textbf{r})$ 
is computed on the grid; this is done by vertically shifting the exact result with the corresponding $\Delta$ and then smoothly switching to the TF result at densities below $\rho'$, using the density-based switching function $f[\rho; \textbf{r}]$ in Eq.~\ref{switch}.  Finally, the NAKP is calculated 
from the smoothly switched kinetic potentials using Eq.~\ref{NAKP}.  The vertical shifts that are applied to kinetic potentials 
simply give rise to an additive constant in the final NAKP, which has no physical effect.  
Although we find that switching to the TF functional at low densities is both convenient and accurate, the protocol described above could be performed using any approximate kinetic energy functional.

\section{Results: Small Systems}

\subsection{Calculation Details}

In this section, e-DFT calculations are presented for the dissociation curves of (H$_2$O)$_2$ and the covalently bound Li$^+$-Be and CH$_3$-CF$_3$ molecules; standard KS-DFT calculations are included for comparison.
All results are obtained using the Molpro quantum chemistry package,\cite{Molpro} with KS-DFT available in the standard version and with the e-DFT method implemented in our modified version.
In the e-DFT calculations, the NAKP is described using either the EE method or the approximate TF\cite{Tho27, Fer28} and LC94\cite{Lem94} kinetic energy functionals; these approaches will hereafter be referred to as e-DFT-EE, e-DFT-TF, and e-DFT-LC, respectively.

All calculations in this section are performed using the B88-P86 exchange-correlation (XC) functional.\cite{Bec88,Per86} Both the XC functional and the NAKP are evaluated on a grid of % the same numerical grid, which is composed of 
Becke-Voronoi\cite{Lam93} cells with resolution to limit the integration error of Slater exchange to 10$^{-12}$ Hartree; the grid is
generated using the Molpro directive GRID=10$^{-12}$. 

The KSCED in Eqs.~\ref{kscedA}-\ref{kscedB} are initialized from the gas phase density of each subsystem, and the eigensolutions for each set of equations are updated at every iteration. Convergence of these equations is improved with the molecular orbital (MO) shifting and direct inversion of iterative subspace (DIIS) algorithms.\cite{DIIS, DIIS2}   For the water dimer, an MO shift of -0.5 Hartree is employed, whereas a -1.0 Hartree shift is used for Li$^+$-Be and CH$_3$-CF$_3$.  Since the DIIS algorithm leads to slow final convergence,\cite{Sha04} it is discontinued once the root mean squared difference (RMSD) of total density matrix elements changes by less than $5 \times 10^{-4}$ between two successive iterations.  The KSCED equations are deemed converged when the total energy of the system changes by less than 10$^{-6}$ Hartree and the RMSD in the total density matrix is smaller than 10$^{-5}$ between two successive iterations. 

For the ZMP step, extrapolation of the solutions to Eq.~\ref{ZMP1e} is performed using $\lambda=\gamma+ \tau j$, where $j=0, 1,\ldots, 5$.  Unless otherwise noted, calculations for the water dimer and Li$^+$-Be employ $\gamma = 5000$ and $\tau=100$, whereas calculations for CH$_3$-CF$_3$ employ $\gamma = 100$ and $\tau=10$. 
To reach adequate convergence, Eq.~\ref{ZMP1e} is solved in several stages.  Firstly, a coarse solution  is reached by using an  MO shift of $-10^{3}$ Hartree and a value of $\lambda = 100$.  Subsequently, using this coarse solution as a starting point, the Eq.~\ref{ZMP1e} solved using a smaller MO shift of $-84$ Hartree and with $\lambda=\gamma$.  Finally, solution of Eq.~\ref{ZMP1e} for each increasing value of $\lambda$ needed for extrapolation employs the solution for the prior value of $\lambda$ as a starting point. The DIIS algorithm is used throughout.  The orbitals from Eq.~\ref{ZMP1e} are deemed converged when the RMSD in the density matrix was smaller than 10$^{-9}$ between two successive iterations; significantly tighter convergence is needed for the ZMP equations than for the KSCED, to ensure an accurate extrapolation.  

Calculations for the water dimer variously employ the aug-pc-3, aug-pc-2, and aug-pc-1 basis sets,\cite{Jen01} in each case using only the s- and p-type functions for the hydrogen atoms and the s-, p-, and d-type functions for the oxygen atoms.  These water dimer basis sets are hereafter referred to as the modified aug-pc-3, aug-pc-2, and aug-pc-1 basis sets, respectively.   
Calculations involving Li$^+$-Be use the s-, p-, and d-type functions of the combined aug-pc-4 and cc-pVQZ (core/valence) basis sets.\cite{Dun95}
In calculations for CH$_3$-CF$_3$, the C atoms are described using the s-, p-, and d-type functions of the combined aug-pc-4 and cc-pV6Z (core/valence) basis sets,\cite{Dun95}
and the H and F atoms  are described using the full aug-pc-1 basis set.\cite{Jen01} Sensitivity of the e-DFT calculations to the basis set is discussed in the next section.

Larger basis sets provide a better description of low-density regions, allowing for the use of smaller values for the parameter $\rho'$ in Eqs.~\ref{switch} and \ref{weights} and providing robustness with respect to the choice of this parameter.    For the water dimer, calculations using aug-pc-3, aug-pc-2, and aug-pc-1 basis sets employ values of $\rho' = 10^{-5}, 10^{-4}$, and $5 \times 10^{-3}$, respectively. For Li$^+$-Be and CH$_3$-CF$_3$, calculations employ $\rho'$ = $10^{-6}$.
In each case, $\xi=10^{-4}$, and the parameter $\kappa$ in Eqs.~\ref{switch} and \ref{weights} is chosen such that $\kappa \rho' = 10$.

\begin{figure} 
  \begin{center} 
    \includegraphics[angle=0,width=8.5cm]{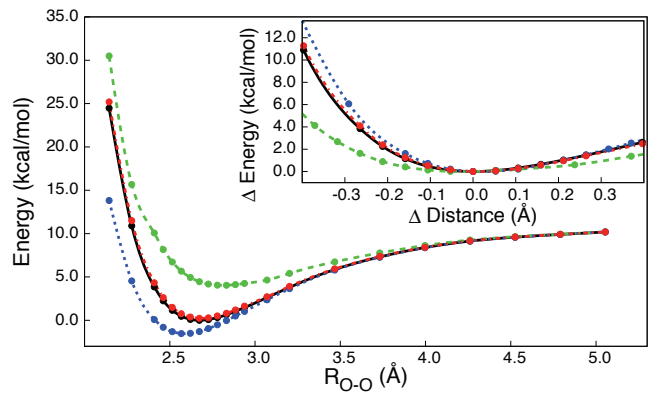}
  \end{center}
  \vspace{-20pt}
\caption{The water dimer dissociation curve, obtained using e-DFT-EE (red, dot-dashed), e-DFT-TF (green, dashed) and e-DFT-LC (blue, dotted).  Also included are reference
reference KS-DFT results (black, solid), which are graphically indistinguishable from the e-DFT-EE results. 
Total energies are plotted with respect to the KS-DFT minimum of -152.430722 Hartree.  Inset, the curves are shifted vertically to align the energy minima and horizontally to align the equilibrium distances.}
\label{fig:waterdimer}
% \vspace{-15pt}
\end{figure}

\subsection{Water Dimer}

Fig.~\ref{fig:waterdimer} presents the dissociation curve for the water dimer, a system with a strong hydrogen bond and significantly overlapping subsystem densities.  The curve is obtained using e-DFT-EE (dot-dashed), e-DFT-TF (dashed), and e-DFT-LC (dotted); KS-DFT results (solid) are also included for reference.  The e-DFT calculations are performed using two embedded subsystems, each corresponding to a different molecule in the dimer.  All calculations presented in the figure utilize the modified aug-pc-3 basis set, with the e-DFT calculations employing the supermolecular basis set convention.  The dissociation curve is plotted as a function of the oxygen-oxygen distance, with the equilibrium water dimer geometry obtained from a KS-DFT energy minimization and with other geometries obtained by displacing the two molecules along the oxygen-oxygen vector while fixing all other internal coordinates.

The e-DFT-EE results in Fig.~\ref{fig:waterdimer} agree well with KS-DFT throughout the range of dissociation distances.  
Numerical results for the two methods are graphically indistinguishable, and the calculated total energies differ by less than 0.5 kcal/mol throughout the entire attractive branch of the curve. 
 Exact numerical agreement between the e-DFT-EE and KS-DFT descriptions is expected only in the limits of an exact XC functional and a complete basis set. 

The  sensitivity of the e-DFT results to approximations in the NAKP is clearly demonstrated in Fig.~\ref{fig:waterdimer}.
 The curve obtained using e-DFT-TF differs significantly from the KS-DFT reference, exhibiting a dissociation energy that is underestimated by 40\% ($\sim$4 kcal/mol) and an equilibrium bond length that is 0.15 {\AA} too long.  Calculations obtained using e-DFT-LC are somewhat improved, although the dissociation energy is still overestimated by 20\% ($\sim$2 kcal/mol) and the equilibrium bond length is underestimated by 0.10 {\AA}. 
 In the inset of Fig.~\ref{fig:waterdimer}, the curvature of the potential energy surfaces in the vicinity of the minimum are compared, revealing significant deviations of the results obtained using the approximate NAKP treatments (e-DFT-TF and e-DFT-LC) with respect to the results obtained using KS-DFT and e-DFT-EE.
 
Iannuzzi and coworkers\cite{Ian06} have demonstrated that e-DFT calculations using approximate treatments of the NAKP, including the TF and LC94 functionals, lead to qualitative failure in describing the structure of liquid water.
Fig.~\ref{fig:waterdimer} illustrates the origin of this failure in terms of the pairwise interactions among molecules, and it suggests that e-DFT-EE  will enable the accurate, first-principles simulation of liquid water and aqueous solutions. Critical to this effort, however, is the efficient and parallelizable implementation of the EE method for large systems, which is discussed in Section V.

\begin{figure} 
  \begin{center}
    \includegraphics[angle=0,width=8.5cm]{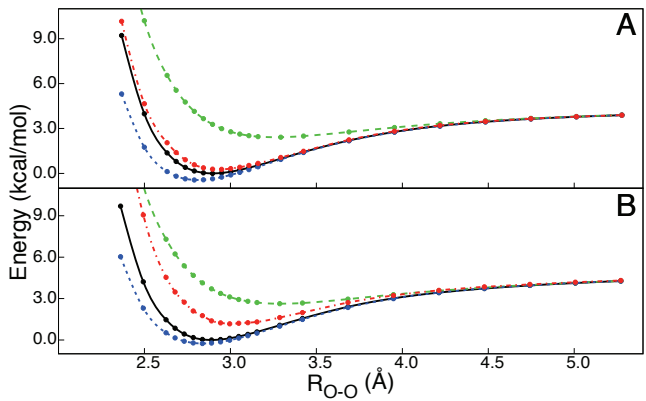}
  \end{center}
  \vspace{-20pt}
\caption{Basis set dependence of the water dimer dissociation curve, illustrated for calculations using the {\bf (A)} modified aug-pc-2 and {\bf (B)} modified aug-pc-1 basis sets.
Results for the e-DFT-EE, e-DFT-TF, e-DFT-LC, and KS-DFT methods are reported as in Fig.~\ref{fig:waterdimer}.
Total energies are plotted with respect to the KS-DFT minimum energies of  -152.953947 Hartree (panel A) and -152.864441 Hartree (panel B).}
\label{fig:waterdimerB}
% \vspace{-15pt}
\end{figure}

The sensitivity of the e-DFT calculations to basis set completeness is illustrated in Fig.~\ref{fig:waterdimerB}, in which the water dimer dissociation curves are recalculated using the modified aug-pc-2 (Fig.~\ref{fig:waterdimerB}A) and modified aug-pc-1 basis sets (Fig.~\ref{fig:waterdimerB}B). 
Comparison of the KS-DFT results and the e-DFT-EE results reveals that the agreement between the methods worsens with smaller basis set; of course, both the KS-DFT calculations and the e-DFT-EE calculations are basis-set dependent.  In the e-DFT-EE calculations, smaller basis sets give rise to numerical artifacts including the oscillatory behavior in the King-Handy expression for the kinetic potential.\cite{Kin00} %and slow convergence in the iterative solution of Eq. \ref{ZMP1e} 
For the modified aug-pc-1 basis set (Fig.~\ref{fig:waterdimerB}B), the reasonable agreement between KS-DFT and e-DFT-LC is due to a fortuitous cancellation of errors from the approximate NAKP functional and the small basis set.

\subsection{Li$^+$-Be}

\begin{figure} 
  \begin{center}
%    \hspace*{-20pt}  \includegraphics[angle=0,width=8.5cm]{figs/BeLi.eps} 
        \hspace*{-20pt}  \includegraphics[angle=0,width=8.5cm]{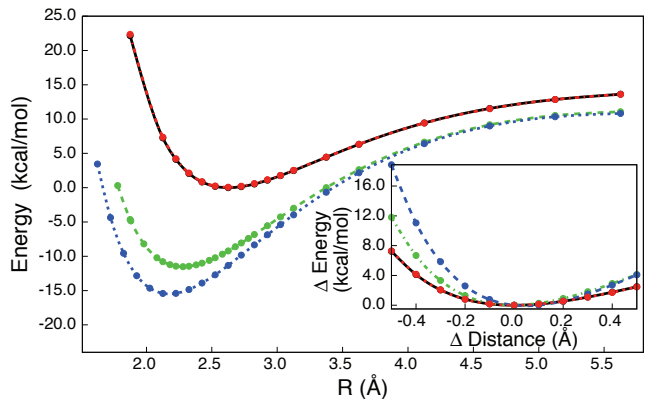} 
  \end{center}
  \vspace{-20pt}
\caption{The Li$^+$-Be dissociation curve.
Results for the e-DFT-EE, e-DFT-TF, e-DFT-LC, and KS-DFT methods are reported as in Fig.~\ref{fig:waterdimer}. 
 The results for e-DFT-EE and the reference KS-DFT results are graphically indistinguishable. Total energies are plotted with respect to the KS-DFT minimum energy of -21.962072 Hartree. Inset, the curves are aligned as in the inset of Fig.~\ref{fig:waterdimer}.
}
\label{fig:LiBe}
\end{figure}

We now consider the heterolytic cleavage of a weak covalent bond,  Li$^+$-Be$\rightarrow$Li$^+$+Be,
using KS-DFT and e-DFT.  The e-DFT calculations were performed in the supermolecular basis set convention using two embedded subsystems, one corresponding to the 2-electron Li ion and the other corresponding to the 4-electron Be atom.
The dissociation curve for  Li$^+$-Be is plotted in Fig.~\ref{fig:LiBe}.

  As is seen from the main figure, the e-DFT-EE calculations accurately reproduce the total energies from KS-DFT throughout the entire range of internuclear distances.  The dissociation curves for these two methods,  which are graphically indistinguishable in Fig.~\ref{fig:LiBe}, deviate by less than 0.2 kcal/mol throughout the range of separations and the dissociation energy deviates by only $0.07$ kcal/mol. 
In contrast, the e-DFT-TF results are in qualitative disagreement with the KS-DFT reference calculations; in addition to dramatically overestimating the dissociation energy of the molecule by approximately 12.5 kcal/mol, the method predicts the equilibrium bond length to be 20\% too short.  Interestingly, the e-DFT-LC method performs significantly worse in this application.  The calculations based on the approximate LC94 kinetic energy functional overestimate the dissociation energy by approximately 16 kcal/mol and predict the equilibrium bond length to be 25\% too short.
The inset to Fig.~\ref{fig:LiBe} illustrates that both e-DFT methods that use approximate treatments for the NAKP lead to an overestimation of the energy surface curvature in the vicinity of the equilibrium bond distance.

The results in Fig.~\ref{fig:LiBe} illustrate the well-known breakdown of e-DFT with approximate treatments of the NAKP for applications involving strongly overlapping subsystem densities.  They further show that our EE method overcomes this large error, yielding the first numerical demonstration of an e-DFT method to describe covalent bond-breaking with chemical accuracy.  Since e-DFT-EE is a formally exact method, 
this result is expected.  However demonstration that the level of accuracy in Fig.~\ref{fig:LiBe} can be achieved in practical numerical simulations constitutes a non-trivial validation of the method. 

\subsection{CH$_3$-CF$_3$}

\begin{figure} 
  \begin{center}
  \includegraphics[angle=0,width=8.5cm]{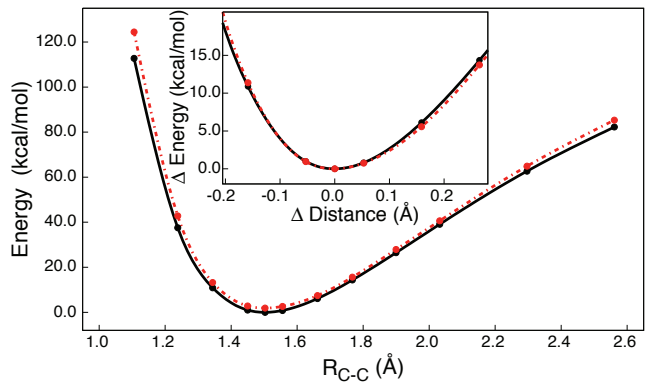} 
   \end{center}
  \vspace{-20pt}
\caption{The CH$_3$-CF$_3$ dissociation curve for heterolytic cleavage of the C-C bond.  Results are presented for the e-DFT-EE (red, dot-dashed) and KS-DFT (black, solid) methods. Total energies are plotted with respect to the KS-DFT minimum energy of -377.575687 Hartree.    Inset, the curves are aligned as in the inset of Fig.~\ref{fig:waterdimer}.  
}
\label{fig:ch3cf3}
\end{figure}

In a more challenging application for e-DFT, we consider the heterolytic cleavage of a strong carbon-carbon $\sigma$-bond, CH$_3$-CF$_3$ $\rightarrow$ CH$_3^+$ + CF$_3^-$.
The e-DFT calculations were again performed in the supermolecular basis set convention using two embedded subsystems, one corresponding to the 8-electron CH$_3^+$ moiety and the other corresponding to the 34-electron CF$_3^-$ moiety.  The geometry for the lowest energy point along the curve is provided in the supplemental information; the dissociation curve in Fig.~\ref{fig:ch3cf3} is plotted by extending the C-C distance while keeping all other internal coordinates unchanged.

The dissociation curves in  Fig.~\ref{fig:ch3cf3} are presented only for e-DFT-EE and the reference KS-DFT calculations.  e-DFT-EE reproduces the KS-DFT reference value for the total energy for the molecule at the equilibrium bond distance to within 1.5 kcal/mol, and the embedding method also recovers the reference value for the equilibrium bond distance.  
Furthermore, as is clear from the inset, e-DFT-EE accurately reproduces the curvature of the energy surface in the vicinity of the equilibrium bond distance.
In contrast, the e-DFT-TF and e-DFT-LC descriptions for this system fail dramatically, predicting total energies at the equilibrium bond distance that deviate from the KS-DFT reference by approximately 730 kcal/mol and 980 kcal/mol, respectively.  For calculations with such strongly interacting subsystems, the failure of e-DFT with approximate descriptions for the NAKP methods  has been previously observed.\cite{Vis09} However, the results for e-DFT-EE in Fig.~\ref{fig:ch3cf3} demonstrate significant progress in the accurate description of covalently interacting subsystems using e-DFT. 

\section{Results: Extension to Larger Systems}
\label{sec:larger}

\subsection{Pairwise treatment of the NAKP}

In the previously described implementation of e-DFT-EE,  the ZMP step, or an equivalent LCS, is performed on the full system of interest.
However, numerical challenges limit the LCS to systems with less than 10-15 atoms,\cite{Han98,Spr00,Toz01,Ron08,Ron09,Toz00}
potentially  hindering the applicability of e-DFT-EE in large systems.
To avoid this problem, we demonstrate a pairwise approximation for the NAKP that enables the scalable implementation of e-DFT-EE.

For a system composed of $N_\textrm{sub}$ embedded subsystems, $\{\rho_{\alpha}\}$, the non-additive kinetic energy can be approximated using a pairwise sum,\cite{tfm10jason} such that
\begin{eqnarray}
T^\textrm{nad}_s[\{\rho_{\alpha}\}]&\equiv&T_s[\bar{\rho}]-\sum_{\alpha=1}^{N_\text{sub}}T_s[\rho_{\alpha}]
\label{pairwise}
\\
&\approx&\sum_{\alpha<\beta=1}^{N_\text{sub}}\left(T_s[\rho_{\alpha}+\rho_{\beta}]-T_s[\rho_{\alpha}]-T_s[\rho_{\beta}]\right),\nonumber
\end{eqnarray}
where $\bar{\rho}=\sum_{\alpha=1}^{N_\text{sub}}\rho_{\alpha}$.
The NAKP for a given subsystem $\alpha$ is then
\begin{equation}
 v_{\textrm{nad}}[\rho_{\alpha}, \{\rho_{\alpha}\}; \mathbf{r}]  = \sum_{\beta \neq \alpha}^{N_\text{sub}} 
  ( v^{\alpha\beta}_{T_{\textrm{s}}}(\textbf{r}) -v^{\alpha}_{T_{\textrm{s}}}(\textbf{r}) ).
  \label{pairwiseNAKP}
 \end{equation} 
 
Applying the EE method to this approximation for the NAKP, 
a ZMP step is performed at each iteration of the KSCED to obtain the KS orbitals corresponding to each pair of subsystems densities, $\{\phi_i^{\alpha\beta}\}$.
 Then, using both the subsystem KS orbitals \{$\phi_i^{\alpha}$\} from the KSCED and the subsystem-pair KS orbitals $\{\phi_i^{\alpha\beta}\}$, the  NAKP is evaluated directly from Eqs. \ref{king_handy} and \ref{pairwiseNAKP}.  
In this approach, only the NAKP is assumed to be pairwise additive; all other interactions in the system are treated with full generality.
Since the ZMP step is applied only to the subsystem pairs, this approach is numerically feasible if each subsystem is limited to a relatively small number of atoms, regardless of the total system size. The short-ranged nature of contributions to the non-additive kinetic energy suggests that distance-based cutoffs can be employed within the sum over subsystem pairs.\cite{tfm10jason}

It was emphasized earlier that the converged results of the ZMP step are independent of the choice of external potential, $v_\textrm{ext}({\bf r})$, in Eq.~\ref{ZMP1e}.  In the pairwise implementation of e-DFT-EE for the water trimer in Sec.~V~B, we employ the following external potential for each pair of densities $\rho_{\alpha}$ and $\rho_{\beta}$, 
\begin{eqnarray}
v_\textrm{ext}({\bf r}) &=& v_{\textrm{ne}}(\mathbf{r})+v_\textrm{J} [ \bar\rho; {\bf r}]  + v_\textrm{xc} [ \bar\rho; {\bf r}]   \nonumber \\
&&+\frac{\delta \tilde{T}_{\textrm{s}}[\bar\rho]}{\delta(\rho_\alpha+\rho_\beta)}
-\frac{\delta \tilde{T}_{\textrm{s}}[\rho_\alpha+\rho_\beta]}{\delta(\rho_\alpha+\rho_\beta)},
\end{eqnarray}
where $\tilde{T}_{\textrm{s}}$ indicates the approximate TF functional.
This external potential approximates the KSCED effective potential (Eq.~\ref{vsced_orig}) for the pair of subsystems embedded within the remainder of the full system;
note that the TF functional is used only to regularize the effective potential for the ZMP step; it does not introduce any additional approximation into the e-DFT-EE calculation.
In Sec.~V~C, we use a simple external potential that includes only the electron-nuclear interactions for the subsystem pair.  

The following two sections demonstrate the accuracy of this pairwise implementation of e-DFT-EE (Sec.~V~B) and the efficiency with which it can be implemented in parallel (Sec.~V~C).

\subsection{Water Trimer Application: Testing Pairwise Additivity in the NAKP}

\begin{figure} 
  \begin{center}
  \includegraphics[angle=0,width=8.5cm]{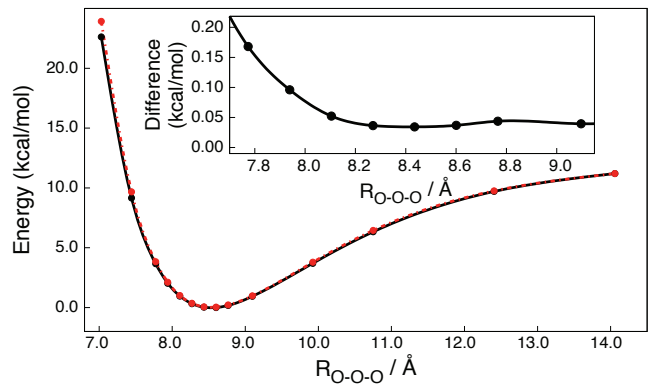}
  \end{center}
  \vspace{-15pt}
\caption{Symmetric dissociation curves for the water trimer, illustrating the pairwise additivity of the NAKP.  
Calculations are performed using the e-DFT-EE method, with no approximation to the NAKP (black, solid) and with the pairwise approximation to NAKP (red, dot-dashed).
The curves are plotted as a function of the sum of the three O-O distances, with details of the molecular geometries provided in the text.
Total energies plotted with respect to the minimum energy of -229.440307 Hartree for the full NAKP treatment.
Inset, the difference between the two curves is plotted.
}
\label{fig:watertrimer}
\end{figure}

Fig.~\ref{fig:watertrimer} presents a test of pairwise additivity in the NAKP (Eq.~\ref{pairwiseNAKP}) for 
a hydrogen-bonded trimer of  water molecules. e-DFT-EE calculations are performed using three embedded subsystems, each corresponding to a different molecule in the trimer.  In a first set of results, the symmetric dissociation curve for the trimer is calculated using no assumptions about the NAKP (solid);
in a second set of results, the curve is calculated using the assumption of pairwise additivity of the NAKP (dot-dashed).
 The equilibrium geometry is provided in the supplemental information; other geometries along the dissociation curve were then obtained by uniformly stretching the oxygen-oxygen distances in the cluster, keeping all other internal coordinates unchanged.
  The trimer calculations were performed using the modified aug-pc-2 basis set with the monomolecular basis set convention; all other calculation details are identical to those described previously for the modified aug-pc-2 calculations of the water dimer.

The agreement between the two curves in Fig.~\ref{fig:watertrimer} indicates that Eqs.~\ref{pairwise} and \ref{pairwiseNAKP} are excellent approximations for the non-additive kinetic energy and NAKP, respectively.   
Throughout the entire attractive branch of the curve the total energies differ by less the 0.5 kcal/mol, and the largest deviations appear only in the strongly repulsive region at short distances.  
 This good agreement is particularly notable, given that the cyclic trimer geometries might be expected to magnify possible non-additive contributions to the total energy; even better adherence of the NAKP to pairwise additivity is expected for linear geometries of the trimer.
We have previously noted that
higher-order corrections to Eqs.~\ref{pairwise} and \ref{pairwiseNAKP} are possible,\cite{tfm10jason} although the results in Fig.~\ref{fig:watertrimer} suggest that the assumption of pairwise additivity will be adequate in many cases. 

\subsection{Parallel Scaling of e-DFT-EE}

Primary bottlenecks in KS-DFT include calculation of the two-electron integrals and solution of the eigenvalue problem.  In standard implementations,
the  two-electron integral calculations scales as $M^4$ and the eigenvalue calculation scales at best as $M^2$,
where $M$ is the total number of basis functions.\cite{Del08,Gu95}  
More efficient methods for computing the two-electron integrals include prescreening,\cite{Sch99b} Ewald summations,\cite{Ham94} and the fast-multipole method;\cite{TeV01} however, solution of the eigenvalue problem remains a computational bottleneck in most KS-DFT implementations.\cite{Saa10}

 As has been noted in previous work,\cite{Ian06} the monomolecular basis set convention leads to advantageous scaling properties for e-DFT.
 In this convention, the number of basis functions used to solve each KSCED, $M_{\textrm{sub}}$, is independent of system size. Consequently,  the
 total cost of the eigenvalue problem scales linearly with the number of subsystems, $N_{\textrm{sub}}$, and it can be trivially parallelized to the subsystem level.

The cost of the two-electron integral calculation is also reduced in the monomolecular basis set convention.  Terms arising from orbitals centered on molecules in more than two different subsystems are exactly zero, such that the total cost of this operation scales with $N_{\textrm{sub}}^2 M_{\textrm{sub}}^4$.
Furthermore, in this convention, the density for each subsystem is spatially localized, such that short-ranged contributions to the KSCED effective potential, including exchange, correlation, short-ranged electrostatic contributions, and pair-wise contributions to the NAKP, can be truncated at a cutoff distance.  Long-ranged electrostatic contributions to the KSCED effective potential can be efficiently treated using Ewald summations or other standard methods.\cite{Ham94,TeV01}  Setting aside these long-ranged terms for the current demonstration, the use of distance-based cutoffs reduces the scaling of the total two-electron integral calculation to $N_{\textrm{sub}} M_{\textrm{sub}}^4$, which can be parallelized to yield constant wall-clock time scaling with increasing system size.

\begin{figure} 
  \begin{center}
  \includegraphics[angle=0,width=9.2cm]{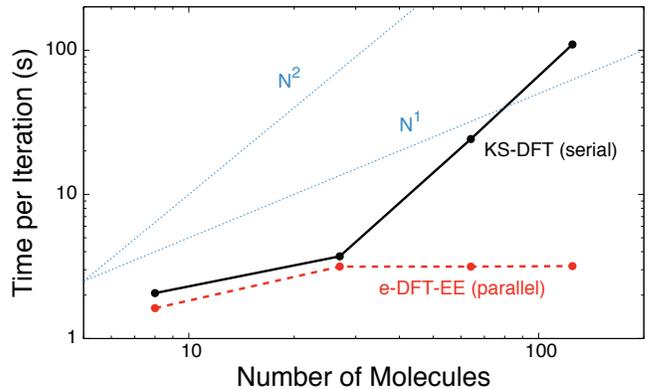} 
  \end{center}
  \vspace{-15pt}
\caption{ Wall-clock timings for lattices of hydrogen molecules, ranging in size from 8 to 125 H$_{2}$ molecules.  The dotted blue lines indicate ideal quadratic and linear scaling, the solid, black curve corresponds to the serial implementation of integral-prescreened KS-DFT in Molpro, and the dashed, red curve corresponds to e-DFT-EE using a number of parallel processors equal to the number of molecules in the system.
}
\label{fig:timings}
\end{figure}

To illustrate these scaling properties, Fig.~\ref{fig:timings} presents
benchmark timings for simple tetragonal lattices of 8 to 125 H$_2$ molecules,
using both e-DFT-EE and the KS-DFT implementation in Molpro. 
The H$_2$ molecules are oriented parallel to the $z$ axis, with a bond length of 
 0.8 \AA, and the centers-of-mass for the molecules are spaced by
 3.0 \AA\ along the $x$ and $y$ axes and by
  3.8 \AA\ along the $z$ axis.  
 All calculations employ the uncontracted STO-3G basis set,\cite{Pop69} Slater exchange\cite{Sla51} without electron correlation, and a grid density that ensures an integration error in the exchange energy of less than $10^{-6}$ Hartree. 
The e-DFT-EE calculations are performed with each molecule defined as a different subsystem, using the monomolecular basis set convention, and using one parallel processor per subsystem.  
Values for the parameters $\lambda$, $\rho'$, $\kappa$, and the MO shift are the same as those used for the Li$^+$-Be system.  
The cutoff for the calculation of the electrostatics, exchange, and NAKP terms is set to 4.0 \AA\ in these calculations, such that only nearest-neighbor molecules in the lattice contribute to these terms.
All calculations are performed on a cluster of dual, quad-core 2.6 GHz Xeon Intel processors with Infiniband communication.

\begin{figure}
  \begin{center}
        \includegraphics[angle=0,width=9.2cm]{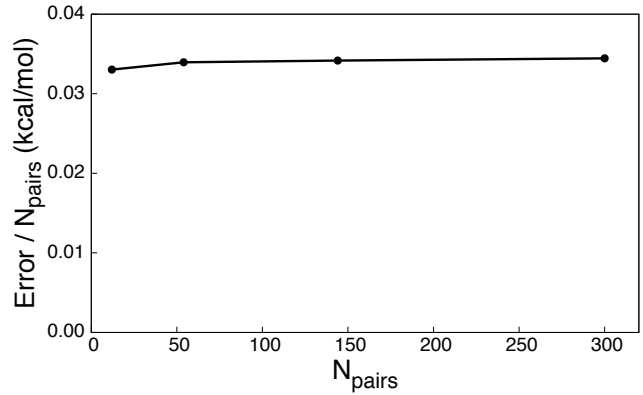} 
  \end{center}
  \vspace{-15pt}
\caption{ Error in the total energy of the e-DFT-EE calculation relative to KS-DFT for increasing system size, plotted with respect to the number of nearest-neighbor pairs.
}
\label{fig:energyerror}
\end{figure}

The timings in Fig.~\ref{fig:timings} indicate that the e-DFT-EE wall-clock time scales independently of the system size, with the deviations at small sizes due the boundaries of the finite crystal.  As expected, the KS-DFT results in the serial Molpro implementation with integral prescreening scales quadratically with the increasing system size.
In Fig.~\ref{fig:energyerror}, relative energy of the e-DFT-EE and the KS-DFT calculations are plotted as a function of the number nearest-neighbor pairs in the lattice, $N_{\textrm{pairs}} = 3(N_{\textrm{sub}}-N^{2/3}_{\textrm{sub}})$.  The error is small and independent of system size.
  The integrated error in the density per molecule was found to behave similarly (not shown).

\section*{CONCLUSIONS}

We introduce a general implementation of the EE method for calculating NAKP contributions in the e-DFT framework, and we present a range of molecular applications. 
The accuracy of e-DFT-EE is demonstrated for systems with covalently bonded and hydrogen-bonded subsystems.  For the dissociation of the water dimer and the covalent bonds in Li$^+$-Be and CH$_3$-CF$_3$, e-DFT-EE preserves excellent agreement with reference KS-DFT calculations, whereas approximate treatments for the NAKP, including those based on the TF or LC94 kinetic energy functionals, lead to known failures.
Furthermore, pairwise approximation of the NAKP yields excellent accuracy for the hydrogen-bonded water trimer, and it enables ideal, constant system-size scaling in applications to molecular clusters with up to hundreds of atoms.
These results establish e-DFT-EE as a promising methodology for performing accurate, first-principles molecular dynamics and for accurately embedding high-level wavefunction methods in complex systems.

\section*{ACKNOWLEDGEMENTS}

This work is supported by the U. S. Army Research Laboratory and the U. S. Army Research Office under grant W911NF-10-1-0202 and by the U. S. Office of Naval Research under grant N00014-10-1-0884. T.A.B.~acknowledges support from a National Defense Science and Engineering Graduate Fellowship, and T.F.M.~acknowledges support from a Camille and Henry Dreyfus Foundation New Faculty Award and an Alfred P. Sloan Foundation Research Fellowship.

\end{document}